\documentclass[aps,showpacs,twocolumn,superscriptaddress]{revtex4}

\usepackage{graphicx}
\usepackage{epsfig}
\usepackage{amsfonts}
\usepackage{amsmath,amssymb}

\begin{document}

\title{Global nuclear structure effects of tensor
  interaction}

\author{M. Zalewski}
\affiliation{Institute of Theoretical Physics, University of Warsaw, ul. Ho\.za
69, 00-681 Warsaw, Poland.}

\author{P. Olbratowski}
\affiliation{Institute of Theoretical Physics, University of Warsaw, ul. Ho\.za
69, 00-681 Warsaw, Poland.}

\author{M. Rafalski}
\affiliation{Institute of Theoretical Physics, University of Warsaw, ul. Ho\.za
69, 00-681 Warsaw, Poland.}

\author{W. Satu{\l}a}
\affiliation{Institute of Theoretical Physics, University of Warsaw, ul. Ho\.za
69, 00-681 Warsaw, Poland.}
\affiliation{KTH (Royal Institute of Technology), AlbaNova University
Center, 106 91 Stockholm, Sweden}

\author{T.R. Werner}
\affiliation{Institute of Theoretical Physics, University of Warsaw, ul. Ho\.za
69, 00-681 Warsaw, Poland.}

\author{R.A. Wyss}
\affiliation{KTH (Royal Institute of Technology), AlbaNova University
Center, 106 91 Stockholm, Sweden}

\date{\today}

\begin{abstract}
A direct fit of the isoscalar spin-orbit (SO)
and both isoscalar and isovector tensor coupling constants to
the $f_{5/2}-f_{7/2}$ SO splittings in $^{40}$Ca, $^{56}$Ni, and $^{48}$Ca
nuclei requires a drastic reduction of the isoscalar SO strength
and strong attractive tensor coupling constants.
The aim of this work is to address further consequences of these strong
attractive tensor and weak SO fields on binding energies, nuclear
deformability, and high-spin states.
In particular, we show that contribution to the nuclear
binding energy due to the tensor field shows generic {\it magic structure\/}
with the {\it tensorial
magic numbers\/} at $N(Z)$=14, 32, 56, or 90 corresponding to the maximum
spin-asymmetries in $1d_{5/2}$, $1f_{7/2}\oplus 2p_{3/2}$,  $1g_{9/2}\oplus
2d_{5/2}$ and $1h_{11/2}\oplus 2f_{7/2}$
single-particle configurations and that these numbers are
smeared out by pairing correlations and deformation effects.
We also examine the consequences of strong attractive tensor fields
and weak SO interaction on nuclear stability at the drip lines, in particular
close to the {\it tensorial doubly magic\/} nuclei and
discuss the possibility of an entirely new tensor-force driven deformation
effect.

\end{abstract}

 \maketitle

\section{Introduction}
\label{sec1}

The primary goal of energy density functional (EDF) methods is to
describe ground-state energies of fermion systems, i.e., in nuclear physics
applications -- masses of nuclei.  The existence of an universal functional
describing {\it exactly\/} masses of odd, odd-odd and even-even nuclei
is warranted by the Hohenberg-Kohn~\cite{[Hoh64]} and
Kohn-Sham~\cite{[Koh65a]} (HKS) theorems.  The HKS theorems, however, provide
no universal rules underlying the construction of such functional. The complexity
of the functional and our lack of knowledge with respect to the in-medium strong interaction
that governs the structure of finite nuclei make the situation even more difficult.
It does not permit the determination for any
{\it ab initio\/} constraints of the nuclear EDF,
except for dilute neutron systems~\cite{[Bul07]}.  It forces us to use effective
functionals with coupling constants fitted directly to the data.  Hence, a proper
selection of empirical data to be used in the~process of constraining parameters
of the functional becomes, irrespective of the form of functional, the~key issue
for the overall good performance of the nuclear density functional
theory (DFT)~\cite{[Klu09]}.

The typical strategy used to construct the nuclear EDF is to start with either
the finite-range Gogny~\cite{[Gog75]} or zero-range Skyrme~\cite{[Sky56xw]}
effective interactions and construct non-local or local functional, respectively,
by averaging the interaction within the Hartree-Fock (HF) method.  The datasets
used to adjust the free parameters of the theory are dominated by bulk nuclear
matter data and by nuclear binding energies of selected doubly magic nuclei with much
lesser attention paid to the single-particle energies (SPE).

The major reason is the effective mass scaling of the
single-particle (SP) level
density, $g$, in the vicinity of the Fermi energy, $\varepsilon_F$.
In homogeneous nuclear matter the SP level density scales according to
the simple rule: $g(\varepsilon_F) \rightarrow \frac{m}{m^*}g(\varepsilon_F)$.
In finite nuclei the situation is slightly more intricate
mostly due to $\mathbf r$-dependence of the $m^* (\mathbf r)$.  Several
authors~\cite{[Ham76],[Ber80w],[Lit06w]} analyzed the SP level density
scaling arguing that the physical density of SP levels around the Fermi
energy can be restituted only after the inclusion of particle-vibration
coupling, i.e., by going beyond mean-field (MF).  This viewpoint is difficult
to reconcile with the effective EDF theories.  Indeed, these theories should
warrant a~proper value of the effective mass through the fit to empirical data
and readjust other coupling constants to this value of $m^*$.  This should
lead to fairly  $m^*$ independent predictions, provided that the (spherical)
SPE are calculated from the differences between the binding energies in
even-even doubly-magic cores and the lowest SP states in odd-$A$
single-particle/hole neighbors.
Within the EDF approach, the mean-field or, more precisely,
Kohn-Sham SP energies computed in e-e doubly magic core serve
only as auxiliary quantities.

A new strategy for fitting the spin-orbit (SO) and tensor
parts of the nuclear EDF was recently suggested by our
group~\cite{[Zal08]}. Instead of performing
large-scale fits to binding energies we proposed a simple and intuitive three-step
procedure that can be used to fit the isoscalar strength of the
SO interaction as well as the isoscalar and isovector strengths of the
tensor interaction.  The entire idea is based on the observation that the
$f_{7/2} - f_{5/2}$ SO splittings in spin-saturated isoscalar $^{40}$Ca,
spin-unsaturated isoscalar $^{56}$Ni, and spin-unsaturated isovector $^{48}$Ca
form distinct pattern that can neither be understood  nor reproduced based
solely on the conventional SO interaction. Following the general philosophy
of the nuclear DFT we compute the $f_{7/2} - f_{5/2}$ SO splittings
from the differences between the binding energies of these doubly-magic cores
and their odd-$A$ neighbors. However, we
use the same functional to calculate both the ground state energies, what is
well justified, as well as the low-lying SP excitations in the odd-$A$
neighbors. How reasonable is the latter assumption it remains to be studied.

The procedure reveals the need for a sizable reduction
(from $\sim$20\% up to $\sim$35\% depending on the parameterization and, in
particular, on the value of $m^*$)
of the SO strength and at the same time, for
much stronger tensor fields as compared to
the commonly used values. The new parameterization
systematically improves the performance of the functional
with respect to SP properties like the SO splittings or the magic
gaps but deteriorates the binding energies~\cite{[Zal08]}.
The aim of the present work
is to address further consequences of strong attractive tensor
and weak SO fields on binding energies, time-even and time-odd polarization
effects, and the nuclear deformability.
The paper is organized as follows. In Sect.~\ref{sec2} we
briefly present the theoretical background of our model.
In Sect.~\ref{sec3}, we show that the contribution to the
binding energy due to the tensor interaction forms a generic pattern closely
resembling that of the shell-correction with the {\it tensorial magic
numbers\/} shifted up as compared to the standard magic numbers toward
$N(Z)$=14, 32, 56, or 90.  The tensorial magic numbers reflect the
maximum spin-asymmetry in $1d_{5/2}$, $1f_{7/2}\oplus 2p_{3/2}$,
$1g_{9/2}\oplus 2d_{5/2}$,  and $1h_{11/2}\oplus 2f_{7/2}$ configurations,
respectively, in the extreme SP scenario at spherical shape.  The
tensorial magic structure is  smeared out by configuration mixing caused by
pairing and deformation effects.  In Sect.~\ref{sec4} we demonstrate
that one can construct the EDF capable to reproduce
reasonably well both the SO splittings as well as binding energies of the
doubly magic spherical nuclei.  In Sect.~\ref{sec5} we discuss the influence
of strong tensor fields on time-even and time-odd polarization effects in
the $f_{7/2} - f_{5/2}$ SO splittings.  In Sect.~\ref{sec6} we
analyze deformation properties of the
new functionals and discuss a possible novel mechanisms related to the onset
of nuclear deformation in the presence of strong attractive tensor fields.
The paper is concluded in Sect.~\ref{sec7}. This analysis complements
our preliminary results communicated in two earlier conference
publications~\cite{[Zal09],[Sat09]}.

\section{Theory: From  two-body
spin-orbit and tensor interactions to energy density functionals and
mean-fields}
\label{sec2}

In our study we explore the local energy density functional
${\mathcal H}({\mathbf r})$.  It is the sum of the kinetic energy and the
isoscalar ($t=0$) and isovector ($t=1$) potential energy terms:
   \begin{equation}\label{eq108}
   {\mathcal H}({\mathbf r}) = \frac{\hbar^2}{2m}\tau_0
               + \sum_{t=0,1} \biggl\{ {\mathcal H}_t({\mathbf
               r})^{\text{even}} + {\mathcal H}_t ({\mathbf r})^{\text{odd}}
               \biggr\} ,
   \end{equation}
which are conventionally decomposed into parts build of bilinear
forms of either only time-even or only time-odd densities, currents
and their derivatives:
\begin{eqnarray}
\label{hte}
\mathcal{H}_t^{\text{even}}
& = & C^{\rho}_t[\rho_0] \rho^2_t + C^{\Delta \rho}_t
\rho_t\Delta\rho_t + \\ \nonumber
&\quad & C^{\tau}_t
\rho_t\tau_t + C^J_t {\mathbb J}^2_t +
C^{\nabla J}_t \rho_t {\mathbf \nabla}\cdot{\mathbf  J}_t,
\end{eqnarray}
\begin{eqnarray}
\label{hto}
\mathcal{H}_t^{\text{odd}} & = &C^{s}_t [\rho_0 ] {\mathbf s}^2_t
+ C^{\Delta s}_t {\mathbf s}_t\cdot\Delta {\mathbf s}_t +  \\ \nonumber
&\quad &C^{T}_t{\mathbf s}_t \cdot {\mathbf T}_t +
C^j_t {\mathbf j^2_t} +
C^{\nabla j}_t {\mathbf s}_t \cdot ({\mathbf \nabla}\times {\mathbf j}_t).
\end{eqnarray}
For the time-even, $\rho_t$, $\tau_t$, and ${\mathbb J}_t$,  and
time-odd, ${\mathbf s}_t$, ${\mathbf T}_t$, and ${\mathbf j}_t$, local
densities we follow the convention introduced in Ref.~\cite{[Eng75]}
(see also Refs.~\cite{[Ben03],[Per04]} and references cited therein).

\vspace{0.3cm}

In the present study we focus on the spin-orbit and
tensor parts of the EDF:
\begin{eqnarray}
\label{htta}
\mathcal{H}^T    & = & C^J_0 {\mathbb J}^2_0 +
                       C^J_1 {\mathbb J}^2_1 , \\
\label{httb}
\mathcal{H}^{SO} & = & C^{\nabla J}_0 \rho_0 {\mathbf \nabla}\cdot{\mathbf J}_0  +
                       C^{\nabla J}_1 \rho_1 {\mathbf \nabla}\cdot{\mathbf
                       J}_1 .
\end{eqnarray}

These two parts of the EDF are strongly tied together through their
mutual and unique contributions to the one-body spin-orbit potential.
This relation can be best visualized by decomposing the
spin-current tensor density ${\mathbb J}_{\mu \nu}$
into scalar $ J^{(0)}$, vector ${J}_\mu$ and symmetric-tensor densities
$J^{(2)}_{\mu \nu}$:
\begin{equation}
\label{spin-current}
{\mathbb J}_{\mu \nu} = \tfrac{1}{3} J^{(0)} \delta_{\mu \nu}
+  \tfrac{1}{2} \varepsilon_{\mu \nu \eta}
{J}_\eta  +    J^{(2)}_{\mu \nu} ,
\end{equation}
\begin{equation}
\label{spin-current2}
{\mathbb J}^2
\equiv \sum_{\mu \nu} {\mathbb J}_{\mu \nu}^2 =
\tfrac{1}{3} (J^{(0)})^2  +  \tfrac{1}{2} {\mathbf J}^2
+ \sum_{\mu \nu  } (J^{(2)}_{\mu \nu})^2 ,
\end{equation}
and going to the spherical-symmetry (SS) limit where
the scalar
$ J^{(0)}$ and the symmetric-tensor densities $J^{(2)}_{\mu \nu}$
vanish identically.  In this limit the spin-current tensor density:
\begin{equation}
{\mathbb J}_{\mu \nu}  \xrightarrow{\text{SS\, limit}}
\tfrac{1}{2} \varepsilon_{\mu \nu \eta} {J}_\eta \quad \text{and} \quad
{\mathbb J}^2 \xrightarrow{\text{SS\, limit}} \tfrac{1}{2} {\mathbf J}^2
\end{equation}
reduces, therefore, to the spin-orbit vector density
having a~single radial component,
${\mathbf J}_t=\frac{{\mathbf r}}{r}J_t(r)$.  The variation of the tensor and
SO parts of the EDF over the radial SO densities
$J(r)$ gives the spherical isoscalar ($t=0$) and isovector ($t=1$) SO
MFs,
\begin{eqnarray}\label{sot} W_t^{SO} &
= & \frac{1}{2r}\left( C^J_t J_t(r) - C^{\nabla J}_t \frac{d\rho_t}{dr}\right)
{\mathbf L} \cdot {\mathbf S},
\end{eqnarray}
which can be easily translated into the neutron ($q=n$) and proton ($q=p$) SO
MFs,
\begin{eqnarray}\label{sotnp} W_q^{SO} &=&
    \frac{1}{4r}\bigg\{(C^J_0-C^J_1) J_0(r) + 2C^J_1 J_q(r)
\\ \nonumber
                  &&-(C^{\nabla J}_0-C^{\nabla J}_1) \frac{d\rho_0}{dr}
                    - 2C^{\nabla J}_1 \frac{d\rho_q}{dr}\bigg\}
{\mathbf L} \cdot {\mathbf S}.
\end{eqnarray}
Below we perform calculations without assuming neither spherical nor
time-reversal symmetries. General expressions
for the SO mean-fields can be found in numerous references, see e.g.
Refs.~\cite{[Eng75],[Per04]}, and will not be repeated here.
We provide here spherical formulas (\ref{sot}) and (\ref{sotnp})
because they are crucial for understanding of our fitting strategy of the tensor
and spin-orbit coupling constants which relies on the $f_{7/2}-f_{5/2}$ SO
splittings in spherical doubly magic $^{40}$Ca, $^{56}$Ni and $^{48}$Ca
nuclei, see Ref.~\cite{[Zal08]} and discussion below.

\vspace{0.5cm}

The functional of the form (\ref{hte})-(\ref{hto}) can be
obtained by averaging the conventional Skyrme effective interaction,
$v_{Sk}({\pmb r})$, within the Skyrme-Hartree-Fock (SHF)
approximation~\cite{[Eng75]}.  In such an approach twenty EDF coupling
constants $C_t$ are uniquely expressed by means of ten auxiliary
Skyrme-force parameters, see~\cite{[Ben03]}.  Hence, the SHF
approximation superimposes relatively strong limitations to the EDF.
It serves, though, as a~reasonable starting point for further
studies.  The approach to the EDF starting from the effective force
indicates also possible ways of generalization of
the nuclear EDF.
In particular, investigation of the tensor component requires
the use of a generalized Skyrme force augmented by
a local tensor interaction $v_{T}({\pmb r})$:
\begin{equation}\label{skyrme}
 v({\pmb r}) =  v_{Sk}({\pmb r}) +  v_{T}({\pmb r})
\end{equation}
where
\begin{eqnarray} \label{v_T}
v_T ( {\mathbf r} )  = \frac{1}{2} t_e \biggl\{
    \bigl[
   3 ( \pmb \sigma_1 \cdot \mathbf k^\prime )
     ( \pmb \sigma_2 \cdot \mathbf k^\prime ) -
     ( \pmb \sigma_1 \cdot \pmb \sigma_2 ) \mathbf k^{\prime\, 2}
                                                 \bigr] \delta ({\mathbf r}) &
                                                     \biggr.  \nonumber  \\
     \biggl.
 +    \delta ( \mathbf r )  \bigl[
  3 ( {\boldsymbol \sigma_1} \cdot \mathbf k )
     ( {\pmb \sigma_2} \cdot \mathbf k ) -
     ( {\pmb \sigma_1} \cdot {\pmb \sigma_2} ) \mathbf k^{ 2}
                                                 \bigr]
                                                            \biggr\}
                                                            \nonumber \\
     + t_o  \bigl[  3 ( \pmb \sigma_1 \cdot \mathbf k^\prime )
  \delta ({\mathbf r})( \pmb \sigma_2 \cdot \mathbf k        )
        -  ( \pmb \sigma_1 \cdot \pmb \sigma_2 )  \mathbf k^\prime
                      \cdot \delta ({\mathbf r}) \mathbf k \bigr],
\end{eqnarray}
and, conventionally,
$\mathbf r  = \mathbf r_1 - \mathbf r_2$ and $ \mathbf k
= - \frac{i}{2} ( \pmb \nabla_1 - \pmb \nabla_2 ) $
are relative coordinate and momentum while
$\mathbf k^\prime$ is complex conjugation of $\mathbf k$
acting on the left-hand side.
By averaging $v_T ( {\mathbf r} )$ within the HF approach, one obtains
the following contribution to the time-even part of the
EDF~\cite{[Per04]}:
\begin{equation} \label{tens_te}
\delta\mathcal{H}_t^{\text{even}}
 =    \tfrac{5}{3} B_t^{T} (J^{(0)}_t)^2
   -  \tfrac{5}{4} B_t^{T} {\mathbf J_t}^2
   +  \tfrac{1}{2} B_t^{T} \sum_{\mu \nu  } (J^{(2)}_{t, \mu \nu})^2
\end{equation}
where
\begin{equation}
\label{B-param}
B_0^T = -\tfrac{1}{8}( t_e + 3t_o )  \quad
B_1^T =  \tfrac{1}{8}( t_e -  t_o ).
\end{equation}
Note, that there are two independent contributions to the tensor
part of the EDF.
The Skyrme force contributes, through the exchange term, to
the tensor part of the EDF in an uniform manner, i.e., it depends
on a~unique coupling constant (\ref{htta}).  In contrast, the tensor force
contributes to the EDF in a~non-uniform way.  Hence, the tensor force generates
a~clear {\it theoretical\/} need to generalize the EDF (\ref{htta}) by
using three independent coupling constants multiplying each of the three
terms appearing in Eq.~(\ref{spin-current2}):
\begin{equation}\label{nonuni}
\mathcal{H}_t^T \longrightarrow
C_t^{J^{(0)}} (J^{(0)}_t)^2  + C_t^{J}   {\mathbf J_t}^2
+ C_t^{J^{(2)}} \sum_{\mu \nu  } (J^{(2)}_{t, \mu \nu})^2 .
\end{equation}
The effects of such an
extension,
where the new coupling constants need to be adjusted
can only be probed
in deformed nuclei.  Since conventional effective interactions and
functionals are rather successful in describing nuclear deformation there
is no first-hand motivation for such a~generalization.  Hence, in the present
study, we do not implement this possible extension of the EDF, and we use
the unique tensor coupling constants $C^J_t$, as defined in Eq.~(\ref{hte}).

\vspace{0.5cm}

The contribution to the
time-odd part of the EDF coming from the tensor interaction is:
\begin{eqnarray}
\delta \mathcal{H}_t^{\text{odd}}
 & = & B_t^{T} \left(
 {\pmb s_t}\cdot {\pmb T_t}  -   3 {\pmb s_t}\cdot {\pmb F_t} \right)
 \nonumber \\
 & + & B_t^{\Delta s} \left( {\pmb s_t}\cdot \Delta {\pmb s_t}
 + 3 ({\pmb \nabla}\cdot {\pmb s_t})^2 \right) \label{tens_to} ,
\end{eqnarray}
where
\begin{equation}
\label{B-odd}
B_0^{\Delta s} =  \tfrac{3}{32}( t_e - t_o )  \quad \text{and} \quad
B_1^{\Delta s} = -\tfrac{1}{32}( 3t_e +  t_o ).
\end{equation}
New terms which appear in the time-odd part of the EDF, namely the $\sim {\pmb
s_t}\cdot {\pmb F_t}$ and $\sim ({\pmb \nabla}\cdot {\pmb s_t})^2$ will not be
considered here mostly due to lack of clear experimental indicators
allowing to fit their strength. Extensive discussion
linking the Skyrme forces to the tensor
component in the EDF including, in particular,
the definition of the density ${\pmb F_t}$ can be found in Ref.~\cite{[Les07]}, see also
Ref.~\cite{[Per04]}.

The starting point of our consideration is always the
conventional Skyrme-force-inspired functional with coupling constants fixed
at the values characteristic for either
SkP~\cite{[Dob84]}, SLy4~\cite{[Cha97fw]}, or SkO~\cite{[Rei99fw]}
Skyrme parametrization.  The variants of the EDF with
tensor and spin-orbit strengths modified along the prescription
of Refs.~\cite{[Zal08],[Sat08]} will be marked
by an additional subscript $T$: SkP$_{T}$, SLy4$_{T}$, and SkO$_{T}$.
In the time-odd sector we test two variants of the functional
with the coupling constants fitted to the empirical values of the $s-$wave
Landau parameters~\cite{[Ost92s],[Ben02afw],[Zdu05xw]}
$g_0$=0.4, $g_0^{\,\prime}$=1.2 and to the Gogny-force values of the $p-$wave
Landau parameters~\cite{[Ben02afw],[Zdu05xw]} $g_1$=--0.19,
$g_1^{\,\prime}$=0.62:
\begin{equation}\label{lanis}
g_0 = N_0 (2C_0^s +
2C_0^T \beta \rho_0^{2/3}), \quad g_1 = -2 N_0 C_0^T \beta \rho_0^{2/3},
\end{equation}
\begin{equation}\label{laniv}
g_0^{\,\prime} = N_0 (2C_1^s + 2C_1^T \beta \rho_0^{2/3}), \quad
g_1^{\,\prime} = -2 N_0 C_1^T \beta \rho_0^{2/3},
\end{equation}
where $\beta = (3\pi^2/2)^{2/3}$, and $N_0^{-1} = \pi^2 \hbar^2 / 2m^\star k_F$
is an effective-mass-dependent normalization factor.
In these variants of the EDF we additionally assume density independence of
$C_t^{s}$ coupling constant, set the spin-surface term $C_t^{\Delta s}$$\equiv$0
to zero, and assume gauge-invariant relations $C_t^j=-C_t^\tau$ and
$C_t^{\nabla j}= C_t^{\nabla J}$.  Concerning the time-odd tensor coupling
constants, $C^T_t$, the following two possibilities will be tested: ({\it i\/})
broken gauge-symmetry scenario with $C^T_t$ fitted to the Landau parameters
and ({\it ii\/}) gauge-invariant scenario with  $C^T_t = - C^J_t$ determined
using the time-even coupling constants $C^J_t$.  The variants of the EDF with
spin fields defined using the Landau parameters will be labeled either by
a~subscript $L_S$ or by a~subscript $L_B$ for the gauge-invariant and the
gauge-symmetry-violating functionals, respectively.

\vspace{0.5cm}

\section{Topology of tensor contribution to the nuclear binding
energy}\label{sec3}

Recent revival of interest in the tensor interaction was triggered by
empirical discoveries of strong and systematic changes in the shell structure
of neutron-rich oxygen~\cite{[Bec06s]}, neon~\cite{[Bel05s]},
sodium~\cite{[Uts04s],[Tri05s]}, magnesium~\cite{[Ney05s]}, or
titanium~\cite{[For04w],[Din05w]} nuclei including new shell gap
opening at $N$=32.  These empirical discoveries were successfully
interpreted within the nuclear shell-model after introducing the so-called
monopole shifts.  To account for the data the monopole shifts are: ({\it
i\/}) attractive between $j^\nu_\gtrless$ and $j^\pi_\lessgtr$ orbitals and
({\it ii\/}) repulsive between $j^\nu_\gtrless$ and $j^\pi_\gtrless$ orbitals
where $j_\gtrless = l\pm 1/2$.  The physical origin of these monopole shifts was
attributed to the tensor
interaction~~\cite{[Ots01],[Ots05],[Hon05]}.
Soon after successful shell-model calculations, the mechanism was confirmed
to work within self-consistent mean-field calculations using finite
range Gogny force augmented by the finite-range tensor interaction,
see~\cite{[Ots06]}.  It was shown in Ref.~\cite{[Ots06]} that, apart from
explaining the shell-structure evolution in light
exotic nuclei, the tensor interaction was also capable to account
for empirical trends in the relative positions of the $1f_{5/2}$ and  $1p_{3/2}$
levels in copper~\cite{[Saw03s],[Korgul]} isotopes
or for the evolution of the single-particle
$1h_{11/2}-1g_{7/2}$ level splittings versus $N$ in antimony~\cite{[Sch04s]}
isotopes.

\begin{figure}[t]
\includegraphics[width=0.4\textwidth, clip]{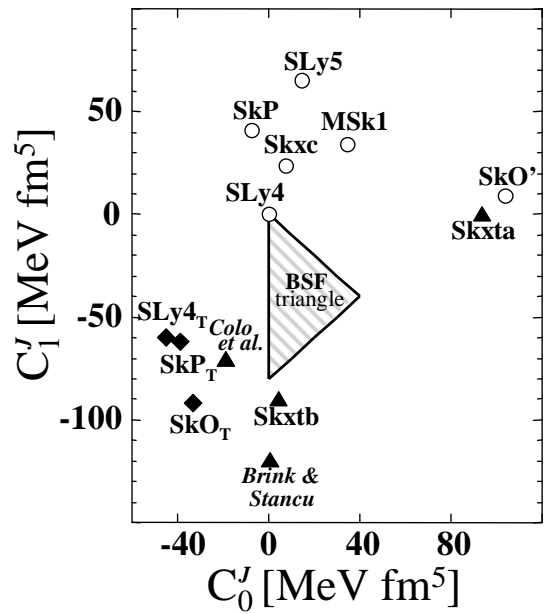}
\caption{The isovector, $C_1^J$, versus
the isoscalar, $C_0^J$, tensor coupling constant.
Open circles mark values representative for several
popular parameterizations fitted predominantly to
the binding energies of spherical nuclei.
Black diamonds represent coupling constants
deduced recently from direct fits to the SPE and the SP
splittings \protect\cite{[Zal08]}. Black triangles represent
fits of Refs.~\cite{[Bro06],[Col07],[Bri07w]}.
The shaded area shows the so-called BSF triangle reflecting the
range of the tensorial parameters deduced in
a~pioneering paper by Brink, Stancu and
Flocard~\protect\cite{[Bri77]}.
}
\label{fig1}
\end{figure}

\begin{figure*}[t]
\includegraphics[width=0.8\textwidth, clip]{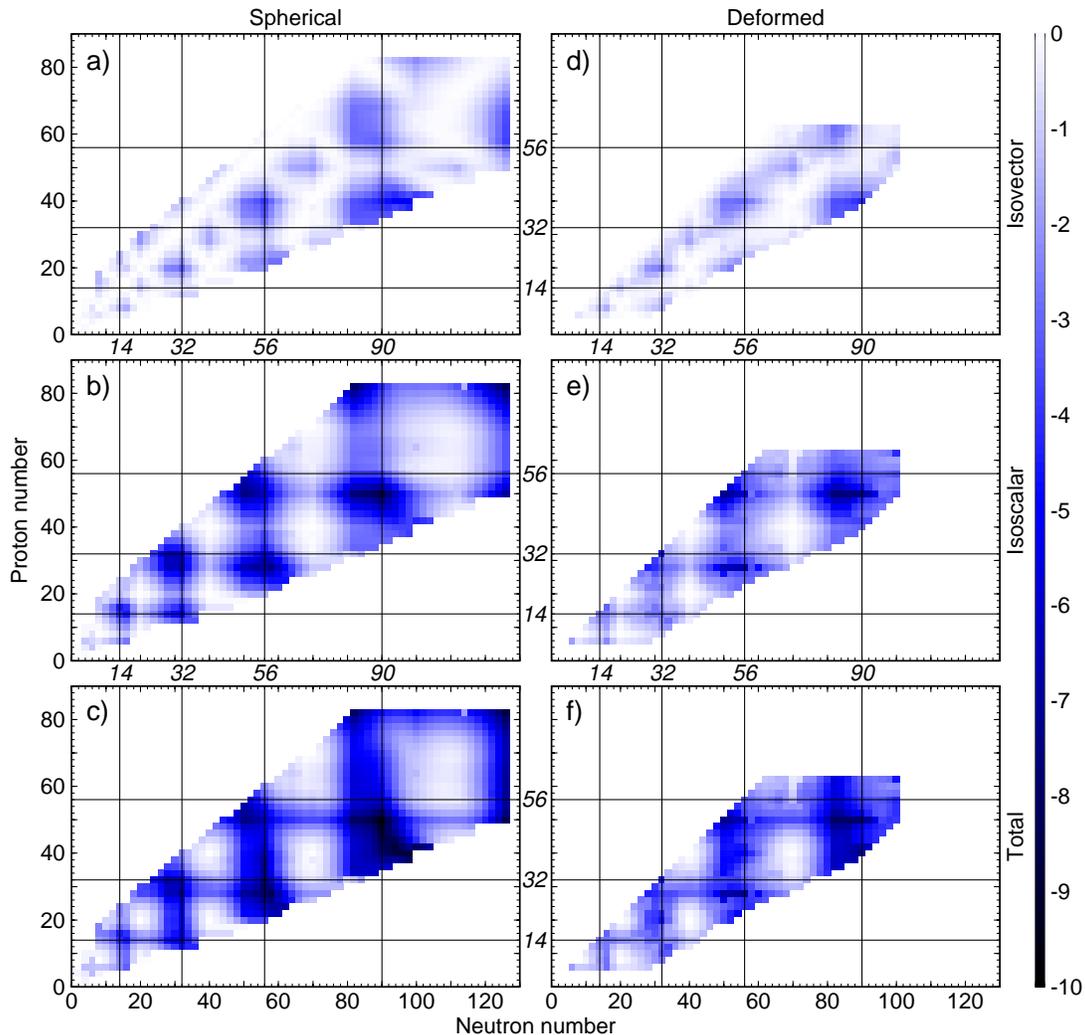}
\caption{(Color online) The isovector (top), the isoscalar (middle) and the total
(bottom) tensor
contribution to the nuclear binding energy obtained from
spherical (left) and deformed (right) HFB calculations.  Both sets of the
calculations were done using the SLy4$_T$ interaction in the particle-hole
channel and the volume-$\delta$ interaction in the particle-particle channel.
Vertical and horizontal lines indicate the single-particle tensorial magic
numbers at spherical shape.  See text for further details.} \label{hfb}
\end{figure*}

\begin{figure*}[t]
\includegraphics[width=0.8\textwidth, clip]{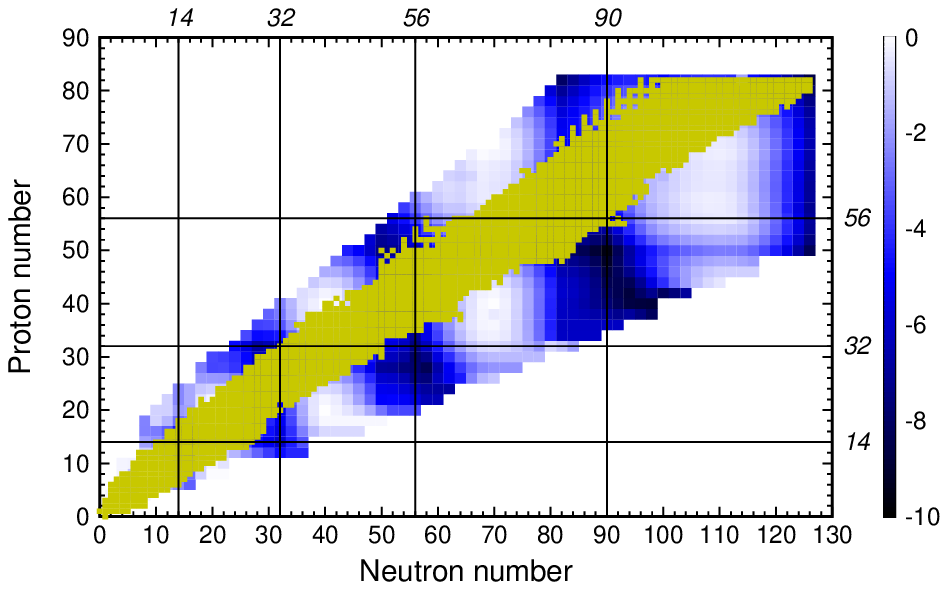}
\caption{(Color online) Tensor contribution to the total binding energy
obtained from spherical HFB calculations.  The map is
overlaid with the map of known, according to Ref.~\protect\cite{[Aud95]},
nuclei.  The figure illustrates these regions where strong tensor
effects may be expected in neutron- and proton-rich nuclei.
}
\label{mapa}
\end{figure*}

The local tensor interaction within the SHF
approximation was studied first in Ref.~\cite{[Bri77]}.
Based on the SPE analysis the effective functional coupling constants
$C_t^J$ were evaluated to lie within the triangle [known as the
Brink-Stancu-Flocard (BSF) triangle] marked schematically in Fig.~\ref{fig1}.
For strictly pragmatic reasons, like technical complexity and lack of
firm experimental indicators  constraining further the BSF estimate,
in many Skyrme parameterizations (including SIII~\cite{[Bei75f]},
SLy4~\cite{[Cha97fw]}, SkM$^*$~\cite{[Bar82f]}, SkO~\cite{[Rei99fw]}),
the tensor terms are simply disregarded by setting $C^J_t\equiv 0$.
Moreover, the $C^J_t$ coupling constants established through fits to,
predominantly, bulk nuclear data seem to contradict the
BSF estimates.  Indeed, the isoscalar tensor coupling constants of
such popular forces like SLy5~\cite{[Cha97fw]}, SkP~\cite{[Dob84]} or
Skxc~\cite{[Bro98]}  are relatively weak while their isovector coupling
constants are positive.  These coupling constants lie outside
the BSF triangle as shown in Fig.~\ref{fig1}.

On the other hand, direct fits to the SP level
splittings~\cite{[Bro06],[Bri07w],[Zal08]} clearly reveal that
drastic changes in the commonly accepted tensor coupling constants are needed
in order to accommodate for the SP data.
This is visualized in Fig.~\ref{fig1} where the new parameterizations are marked
by black triangles and black diamonds, respectively.  The fact that new
values of $C_t^J$ are still slightly scattered is a~consequence of
different fitting strategies as well as different starting
point parameterizations used by different groups.

\vspace{0.3cm}

Our strategy of fitting the coupling constants of the nuclear EDF,
see Ref.~\cite{[Zal08]}, differs
from the strategies applied by other groups.
Unlike the other groups, we fit simultaneously the isoscalar spin-orbit,
$C_0^{\nabla J}$, as well as the isoscalar, $C_0^J$, and
the isovector, $C_1^J$, tensor coupling constants using
a~simple three-step method.  The entire idea of our procedure is based
on the observation that the empirical $1f_{7/2}-1f_{5/2}$ SO splittings in
 $^{40}$Ca, $^{56}$Ni, and $^{48}$Ca form a
distinct pattern, which cannot be reproduced by using solely the
conventional SO interaction.  The readjustment
of the coupling constants goes as follows:
({\it i\/}) in the first step $C_0^{\nabla J}$ is established in the
isoscalar spin-saturated nucleus  $^{40}$Ca; ({\it ii\/}) next,
the $C_0^J$ coupling constant is readjusted in spin-unsaturated isoscalar
nucleus $^{56}$Ni; ({\it iii\/}) finally, the $C_1^J$ coupling constant
is readjusted to the spin-unsaturated isovector nucleus $^{48}$Ca.

Our results, see Refs.~\cite{[Zal08],[Sat08],[Zal09],[Sat09]},
show that drastic changes in the isoscalar SO strength
and the tensor coupling constants are required as compared
to the commonly accepted values.  In turn, one obtains systematic
improvements for such single-particle properties like SO splittings
and magic-gap energies.  It is also interesting to note that
the isoscalar SO and the isoscalar tensor coupling constants
resulting from such a fit are, to large extent, independent on the
parameterization and equal to $C^{\nabla J}_0\approx -60\pm 10$\,MeV\,fm$^5$
and $C_0^J \approx -40\pm 10$\,MeV\,fm$^5$, respectively.  The uncertainties
are rough estimates reflecting the sensitivity of the method.  The isovector tensor
coupling constant, $C^J_1$, is less certain.  It depends on the actual ratio
$C^{\nabla J}_0/C^{\nabla J}_1$ of the SO coupling constants which, in the
adjustment process, was kept fixed to its Skyrme force value.  This is due to
lack of empirical data in $^{48}$Ni which does not allow for firm independent
readjustment of the fourth coupling constant $C^{\nabla J}_1$.

The influence of the tensor interaction on nuclear SPE and SP level
splittings has been analyzed by many authors.  It was shown
that the tensor interaction leaves unique and robust fingerprints
when the SPE and SP level splittings are studied along isotopic
or isotonic chains.  It appears also that the contribution to the binding
energy coming from the tensor interaction, $\delta B_T (N,Z)$, shows several
highly interesting and robust topological features.
In particular, the contribution $\delta B_T (N,Z)$ shows a~generic pattern
closely resembling that of a shell-correction.
The {\it tensorial magic numbers\/} at $N(Z)$=14, 32, 56, or
90 correspond to the maximum spin-asymmetries in $1d_{5/2}$,
$1f_{7/2}\oplus 2p_{3/2}$,  $1g_{9/2}\oplus 2d_{5/2}$ and $1h_{11/2}\oplus
2f_{7/2}$ single-particle configurations, respectively, in the extreme SP
scenario at spherical shape.  The robustness, i.e., model independence of the
tensorial magic pattern, results from rather unambiguously established order of
single-particle levels which is relatively well reproduced by state-of-the-art
nuclear MF models, in particular, in light and medium-mass nuclei.  Note, that
the tensorial magic numbers are only slightly shifted as compared to classic
magic numbers at $N(Z)$=8, 20, 28, 50, and 82.

The tensorial magic pattern is clearly visible in Fig.~\ref{hfb}.
Panels {\bf a)-c)} show  the contribution to the binding
energy coming from:
{\bf a)} the isovector part of the tensor term, $\delta E^T_1 = C_1^J\int d^3
{\pmb r} {\mathbb J}_1^2 ({\pmb r})$, {\bf b)} the isoscalar part of the
tensor term, $\delta E^T_0 = C_0^J\int d^3 {\pmb r} {\mathbb J}_0^2 ({\pmb
r})$, and {\bf c)} the total tensor contributions to the EDF.  These
calculations were performed using spherical Hartree-Fock-Bogolyubov (HFB) code
HFBRAD~\cite{[Ben05]} with the SLy4$_T$ functional of Ref.~\cite{[Zal08]} in the
particle-hole channel and the volume $\delta$-interaction in the pairing
channel.  This part of Fig.~\ref{hfb} shows several interesting features
including:
\begin{itemize}
\item Additional smearing of the SP tensorial magic structure due
to configuration mixing caused by nuclear pairing.  In light and medium-mass
nuclei a substantial tensor contributions located in relatively broad regions
centered around the SP tensorial magic numbers.  In heavier nuclei, where
pairing effects are relatively stronger due to larger density of SP levels,
the erosion of the SP tensorial magic structure is stronger.  The maximum of the
tensor contribution is shifted away from the SP magic numbers. The details, however,
are strongly model-dependent mostly due to large uncertainties in the
positions of the SP levels in heavier nuclei.  Indeed, Skyrme models have
persisting problems to reproduce absolute positions of the experimental SPE as
shown recently in Ref.~\cite{[Kor08]}.

\item
Contribution due to the isovector part of the tensor
interaction is much weaker than the isoscalar contribution.  This
conclusion depends on the tensorial coupling constants.

\item
The isoscalar tensor interaction creates {\it oscillatory\/}
effects in nuclear masses that depend on the degree of spin-unsaturation in a
given nucleus. This additional non-uniform
$N$ and $Z$ dependence may, in particular, obscure conclusions deduced from widely
used binding-energy indicators technique.

\end{itemize}

The second major source of configuration mixing  is due
to the spontaneous breaking of spherical symmetry inherent to the
MF method.  The influence of nuclear deformation on the topology of
the tensor contributions to the binding energy is illustrated in
Fig.~\ref{hfb}d,e,f.  The calculations presented in the figure were
performed for e-e nuclei with $ N \geq Z$ ranging from $6 \leq
Z \leq 64 $ using the HFODD code~\cite{[Dob04fw]}.  The
same SLy4$_T$ interaction of Ref.~\cite{[Zal08]} was used in the p-h channel
and volume-$\delta$ interaction in the p-p channel.
It is clearly visible that the
effect of deformation does not change the topology of the tensor energy
contribution but strongly reduces its magnitude.  One should stress though that
quantitative estimate of the deformation effect is uncertain. The magnitude of
the deformation is extremely sensitive to the balance between SO and tensor
strengths. This effect will be discussed in detail in Sect.~\ref{sec5}.

\vspace{0.3cm}

Fig.~\ref{mapa} shows again the total contribution to the
binding energy calculated using the spherical HFB model.
The map is overlaid with the map of known, according to Ref.~\cite{[Aud95]},
nuclei.  The aim of the figure is to illustrate mass-regions where enhanced
tensor effects, and in turn perhaps new physics, may be expected on
the neutron-rich and proton-rich side.  On the neutron-rich side the regions of
interest, i.e., those which are or can be accessible
experimentally in the nearest future,
include: $Z\approx 14$ and $N\approx 32$, $Z\approx 32$ and $N\approx
56$, and $Z\approx 56$, $N\approx 90$.  In particular,
recent measurements of exotic $^{40}_{12}$Mg$_{28}$ and
$^{42}_{13}$Al$_{29}$ by Bauman
{\it et al.\/}~\cite{[Bau07s]} (see also the discussion in Ref.~\cite{[Hee07]})
approach closely the first of the above
mentioned mass regions.  However, whether or not extra binding due to the
strong attractive tensor interaction gives rise to stabilization of these
nuclei and nuclei around them remains to be studied.  Mean-field
calculations using conventional Skyrme forces predict these nuclei to be
bound~\cite{[Witek]}.

\section{Energy density functional fitted to the single-particle
spin-orbit splittings and to the total binding energies of
spherical nuclei}\label{sec4}

\begin{figure}[t]
\includegraphics[width=0.4\textwidth, clip]{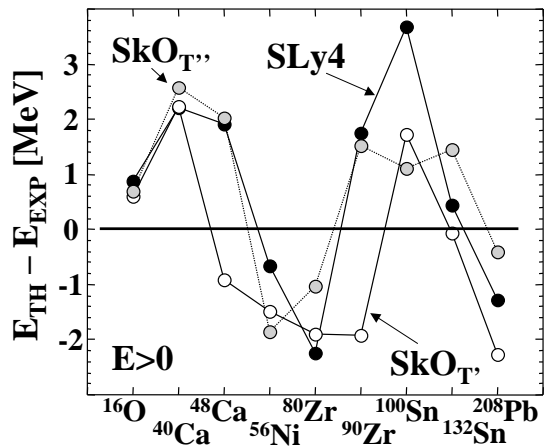}
\caption{Differences between theoretical and experimental binding
energies in spherical doubly magic nuclei and in
$^{80}$Zr.  The calculations have
been done using the SkO$_{T^{\prime}}$ (white dots) and
SkO$_{T^{\prime\prime}}$ (gray dots) functionals.
The conventional SLy4 Skyrme-force result (black dots)
is also shown for the sake of comparison.}
\label{b-ener} \end{figure}

The topology of the tensor contribution to the total binding
energy is, as discussed above, a~generic feature related to
shell-structure and the degree of spin-saturation. The quantitative
features including the total magnitude and the isovector to isoscalar
ratio of the tensor contributions depend, however, upon the actual values
of the tensorial coupling constants.  The two strategies
of fitting effective forces, namely the conventional one based on the
large-scale fit to the binding energies and the one based on the fit of
$C_t^J$ and $C_t^{\nabla J}$ coupling constants directly to the SO
splittings seem to yield contradicting results.  This is clearly
visible in in Fig.~\ref{fig1} where the ranges of the $C_t^J$ strengths
deduced using these two methods simply exclude each other.
% Recently,
%Lesinski {\it et al.}~\cite{[Les07]} have performed systematic adjustments
%of the Skyrme parameters including tensor term to, predominantly,
%bulk nuclear-matter data and binding energies of spherical nuclei.
%They created 36 new parameterizations covering broad range of the
%tensorial coupling constants.  None of these parameterizations
%though is published in their article~\cite{[Les07]}, what signals that
%the authors have rather limited confidence in their quality.
%In fact it looks like they end up with privileging conventional values
%for the spin orbit and tensor interactions, very different from the ones
%suggested, in particular, in our papers Ref.~\cite{[Zal08],[Sat08]}.

\vspace{0.3cm}

A direct fit to the SO splittings leads to drastic changes in the isoscalar SO
strength and the tensor coupling constants as compared to the
commonly accepted values.  In particular, the pronounced reduction of the SO strength
which varies from $ 20$\% for large effective mass,
$m^*$, forces to even $35$\%  for low  ($m^*\approx 0.7$) forces imperils
the agreement with observed binding energies.  This effect cannot
be solely compensated by strong attractive tensor fields.
Hence, further readjustments of the other coupling constants of the EDF
are necessary to assure good performance on masses.

In Ref.~\cite{[Zal08]} we have demonstrated that a considerable improvement
in the quality of mass fits can be achieved be relatively small
readjustments of the EDF coupling constants.
For the purpose of this work we have performed similar calculations
using multi-dimensional minimization technique but starting from
the SkP$_T$ force.  In both cases tiny modifications (of the order of
a~fraction of a percent) in the coupling constants clearly improve the
quality of the mass fit as compared to the SLy4$_T$ and SkO$_T$
forces but still do not provide the quality of the original
SLy4 and SkO parameterizations.

Inherent to the multi-dimensional minimization technique is a~merit
function being a~subject of minimization.  In our calculations
the merit function is constructed out of relative deviations
from measured masses of selected spherical doubly magic nuclei.
It appears that the merit function  in the multi-dimensional space
spanned by the EDF coupling constants varying in the minimization process
is very steep for some specific directions and extremely flat in
others. It
implies that the entire minimization problem is not well defined.  We will
visualize this by taking as a~starting point the SkO parameterization.

By reducing the SO strength by $15$\%  corresponding to
$C_0^J\approx -65.6$\,MeV\,fm$^5$ and  $C_1^J\approx 84.5$\,MeV\,fm$^5$
and taking $C_0^J = -44.1$\,MeV\,fm$^5$
and $C_1^J = -91.6$\,MeV\,fm$^5$ we create a modified version of
the SkO$_T$ parameterization of Ref.~\cite{[Zal08]}.
This parameterization,  dubbed SkO$_{T^\prime}$, has
slightly stronger SO term and more attractive tensor
fields as compared to the SkO$_{T}$.  These changes aim to improve mass
performance of the SkO$_{T}$.  Direct calculations show, see
Fig.~\ref{b-ener}, that the SkO$_{T^\prime}$ functional reproduces masses at a
similar level of accuracy as one of the most popular SLy forces.
The calculations illustrated in Fig.~\ref{b-ener} were performed using the
code HFODD with 20 spherical shells.  We were forced to use the HFODD code
because the spherical HFBRAD code has no two-body center-of-mass correction
implemented.

The performance of the SkO$_{T^\prime}$ force can be further improved
in many different ways.  One example, dubbed
SkO$_{T^{\prime\prime}}$, is illustrated in Fig.~\ref{b-ener}.  This
force was obtained by readjusting isoscalar and isovector central fields
in the following way: $C_0^\rho \rightarrow 1.00015C_0^\rho$
and $C_1^\rho \rightarrow 0.99C_1^\rho$.  As a~result, standard deviation
drops from $\sigma (\text{SkO}_{T^\prime}) \approx 1.663$\,MeV  to
$\sigma (\text{SkO}_{T^{\prime\prime}}) \approx 1.475$\,MeV where, for
comparison, $\sigma (\text{SLy4}) \approx 1.879$\,MeV.  Similar improvements can
be made by, for example, readjusting the density dependent term.

A reasonable performance of the
SkO$_{T^\prime}$ or SkO$_{T^{\prime\prime}}$
functionals with respect to the binding  energies of spherical
doubly-magic nuclei is of great interest. It may help to resolve the
conflict concerning the tensor and SO coupling constants
resulting from (local) fits to the SO splittings and
the SPE~\cite{[Zal08],[Kor08]} preferring
strong tensor and weak SO terms on one hand and from large-scale fits
to the binding energies~\cite{[Les07]} pointing toward
weak tensor and stronger SO terms on the other hand.
Indeed, our present result indicates that
one should explore in large-scale fits functionals
having non-standard forms including, in particular,
functionals having non-conventional
isovector spin-orbit which characterize the SkO functional.

\begin{figure*}[t]
\includegraphics[width=0.8\textwidth, clip]{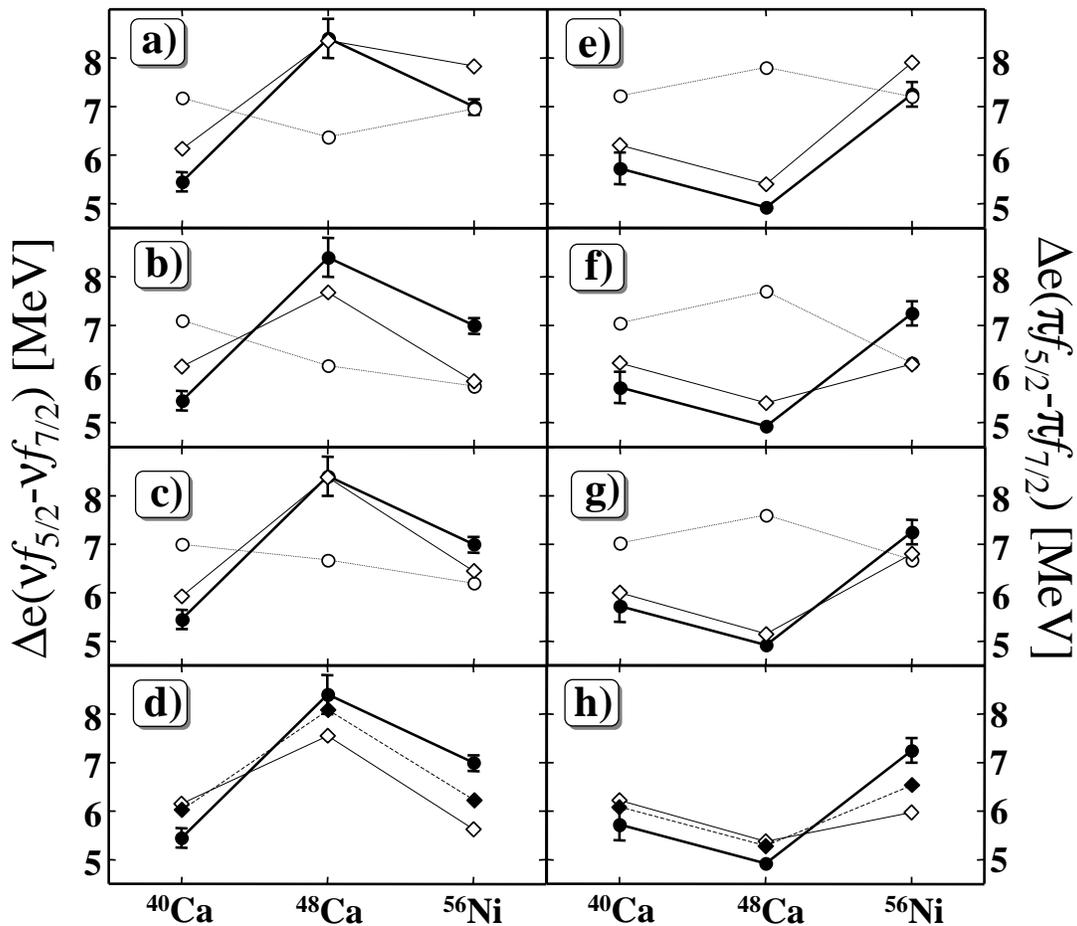}
\caption{SO splittings between the neutron
$\nu 1f_{5/2}$ and $\nu 1f_{7/2}$ (left part) and the proton
$\pi 1f_{5/2}$ and $\pi 1f_{7/2}$ (right part) orbitals
in $^{40}$Ca, $^{48}$Ca, and $^{56}$Ni nuclei.
Filled (open) circles mark empirical and theoretical splittings
calculated using the conventional SkO force, respectively.
Open and filled diamonds indicate calculations performed using
the SkO$_{T^\prime}$ functional.  Panels {\bf a)} and  {\bf e)}
show bare, unpolarized SO splittings deduced directly from the SP spectra
calculated in the doubly magic $^{40}$Ca, $^{48}$Ca, and $^{56}$Ni
nuclei.  Parts {\bf b)} and  {\bf f)}
include only time-even (mass and deformation) polarization effect.
Panels  {\bf c)} and  {\bf g)} include both the time-even and
time-odd polarization effects.
Finally, figures  {\bf d)} and  {\bf h)} show two variants of the
calculations with spin fields readjusted to match empirical
Landau parameters, namely the SkO$_{T^\prime L_S}$ (open diamonds)
and the SkO$_{T^\prime L_B}$ (filled diamonds) functionals.}
\label{polar}
\end{figure*}

\section{Time-even and time-odd polarization effects in the
presence of strong tensor fields}\label{sec5}

The aim of this section is to analyze the polarization
phenomena in the presence of strong tensor fields and to show that,
in spite of the relatively large readjustments as compared to the
SkO$_T$ parameterization, the SkO$_{T^\prime}$ functional can still reproduce
the empirical $1f_{5/2}$ and $1f_{7/2}$ SO splittings quite well.
Fig.~\ref{polar} illustrates the neutron (left hand side) and
proton (right hand side) SO splittings between the $1f_{5/2}$ and $1f_{7/2}$
SO partners in $^{40}$Ca, $^{48}$Ca, and $^{56}$Ni.  The empirical data are
marked by filled circles.  The values shown are average means of empirical results
taken from  Refs.~\cite{[Oro96w],[Sch07a]} (see also
Table III in Ref.~\cite{[Zal08]} for
compilation of the empirical SO splitting data).
Error bars represent deviations from the mean-values.

Open circles illustrate the results of
our calculations using the original SkO parameterization.
Open and filled diamond represent calculations using SkO$_{T^\prime}$
functional.  In all variants of the calculations the $C_t^{\Delta s}$
strength was set to zero to assure convergence.
Different panels represent different variants of the calculations
concerning treatment of the time-odd sector.  Panels {\bf a)} and  {\bf e)}
show bare, unpolarized SO splittings deduced directly from the SP spectra
calculated in the doubly magic $^{40}$Ca, $^{48}$Ca, and $^{56}$Ni
nuclei.  Results presented in the other panels are
calculated from the binding energies
in the doubly magic cores and their one-particle(hole)
odd-$A$ neighbors following the prescription given in
Ref.~\cite{[Zal08]}.  In the calculations
the odd-$A$ binding energies correspond to
fully aligned $\langle I_y \rangle = j$ states at oblate, for one-particle,
and prolate, for one-hole, nuclei, respectively.
Unlike in our previous study~\cite{[Zal08]}, the present calculations include
polarization effects exerted by the odd particle or hole on the even-even core
in the presence of strong attractive tensor fields.

In order to visualize the role of the tensor interaction, in particular
in the time-odd sector, we performed three different variants of the
calculations.  Figures {\bf b)} and  {\bf f)}
include only time-even (mass and deformation) polarization effect.  These
results were obtained by setting all time-odd coupling constants
to zero.  Panels  {\bf c)} and  {\bf g)} illustrate the effect of
the time-odd fields.  These results include both the time-even and
time-odd
polarizations. In these calculations we use gauge invariant functional with
the $C_t^T = - C_t^J$ tensor coupling constants fitted to
the SO splittings.  All other
coupling constants in this run, except $C_t^{\Delta s}\equiv 0$, are
equal to the values given by the SkO parameterization.  Note that the
splittings calculated in this way match almost perfectly the empirical
data.  Note also, that the time-odd polarization effects are indeed
large, reaching a few hundred keV.

Figures  {\bf d)} and  {\bf h)} show two variants of the
calculations with spin fields readjusted to match empirical
Landau parameters, namely the SkO$_{T^\prime L_S}$ and
the SkO$_{T^\prime L_B}$ (see Sect.~\ref{sec2}).
The SkO$_{T^\prime L_S}$ variant is labeled
by open diamonds.  It corresponds to fully gauge
invariant functional.  In this variant we use the $C_t^T = - C_t^J$ tensor
coupling constants fitted to the SO splittings and the spin-field
coupling constants readjusted to the $s$-wave Landau parameters
according to Eqs.~(\ref{lanis})-(\ref{laniv}).

In the SkO$_{T^\prime L_B}$ variant, which is labeled by filled diamonds, we
readjust first the $C_t^T$ coupling constants to the
Gogny values of the $p$-wave Landau parameters $g_1 =-0.19$ and
$g_1^\prime = 1.2$.  This leads to gauge-symmetry violating functional
with the tensorial coupling constant
$C_0^J \approx -44.1$\,MeV\,fm$^5$ and $C_1^J
\approx -91.6$\,MeV\,fm$^5$ in the time-even channel and
 $C_0^T \approx 9.2$\,MeV\,fm$^5$ and $C_1^T \approx -29.9$\,MeV\,fm$^5$
in the time-odd channel which are used
 subsequently to calculate spin-fields coupling constants.
Note, that the SO splittings are quite sensitive to the
way the functional is set up in the time-odd sector.
It means that the entire concept of
fitting the time-odd coupling constants to the Landau parameters in the
presence of the strong tensor terms must be reconsidered.
In particular, the $p$-wave parameters deduced from the
Gogny force $g_1 =-0.19$ and $g_1^\prime
= 1.2$ lead to the $C_t^T$ coupling constants which are completely
inconsistent with the time-even values $C_t^J$ deduced from the
SO splittings.  In turn, the SkO$_{T^\prime L_S}$  and SkO$_{T^\prime L_B}$
have entirely different spin fields with coupling constants equal
$C_0^s \approx 426.4$\,MeV\,fm$^5$ and $C_1^s \approx 48.6$\,MeV\,fm$^5$
and
$C_0^s \approx 18.0$\,MeV\,fm$^5$ and $C_1^s \approx 155.9$\,MeV\,fm$^5$,
respectively.

\section{Effect of tensor field on nuclear deformation} \label{sec6}

\begin{figure*}[t]
\includegraphics[width=0.8\textwidth, clip]{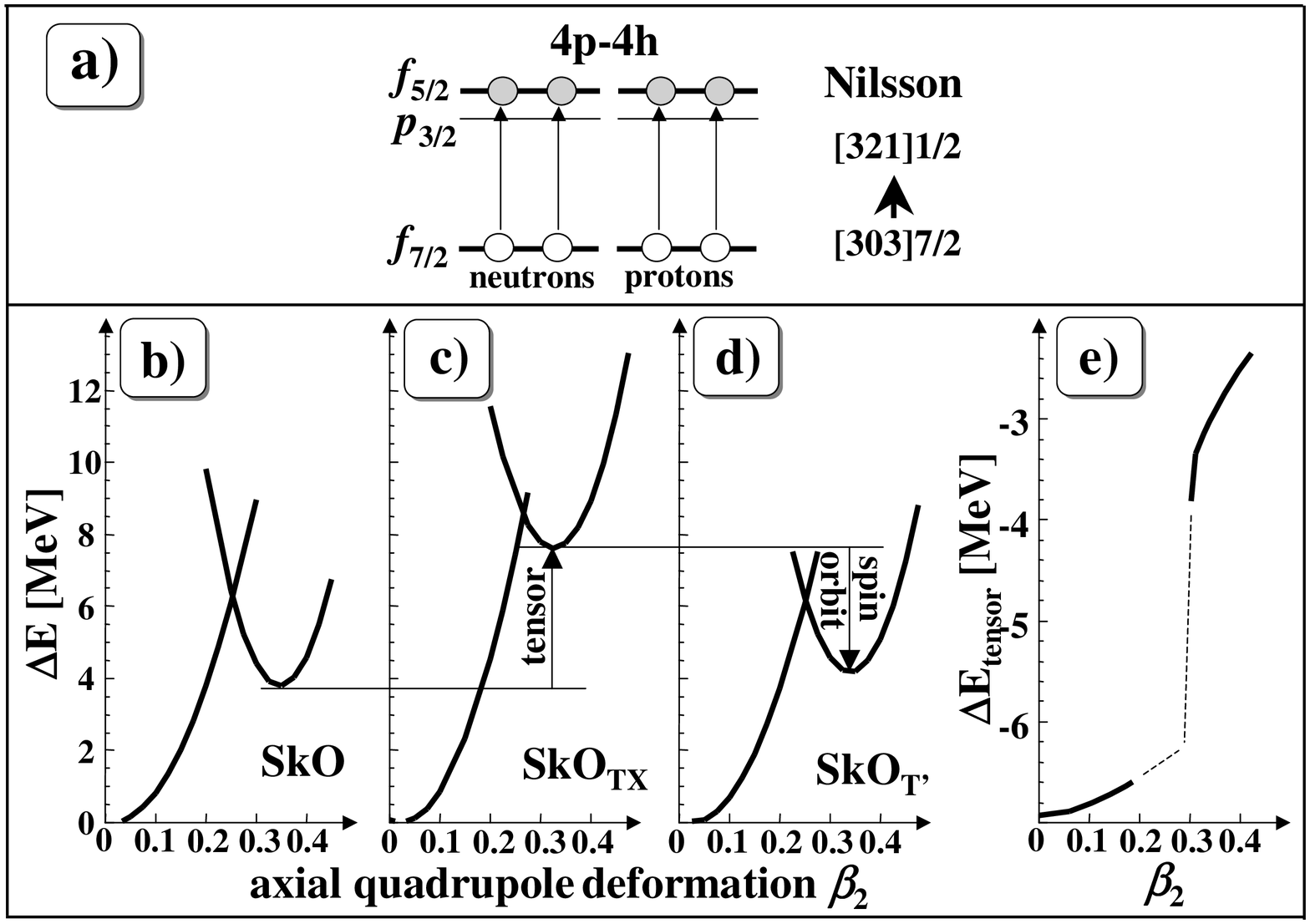}
\caption{Part {\bf a)} shows schematically the~mechanism underlying
the formation of the SD band in
$^{56}$Ni.  It is associated with 4p-4h isoscalar excitation
from the $1f_{7/2}$ to the $1f_{5/2}$ orbital.  Figures {\bf b)}--{\bf d)}
show potential energy curves for the GS and the SD bands calculated using
the SkO, SkO$_{TX}$, and SkO$_{T^\prime}$ functionals,
respectively.   Parts {\bf b)} and {\bf c)}
demonstrate the effect of
strong attractive isoscalar tensor field and  part  {\bf d)}
the effect of reduced spin-orbit field on the excitation
energy of the SD band.
Part {\bf e)} shows the change in the tensor energy associated with
the 4p-4h excitation leading to the SD band.
}
\label{56ni}
\end{figure*}

In the previous section we have shown that one can construct
the EDF capable to reproduce reasonably well binding energies
of the spherical doubly magic nuclei and, simultaneously, account for
the SO splittings of the $1f_{7/2}-1f_{5/2}$ SP levels.
By inspecting closer the results presented in Fig.~2 one observes very strong
time-even (mass and deformation) polarization effect on the calculated SO
splitting in $^{56}$Ni -- compare Fig.~\ref{hfb}a and Fig.~\ref{hfb}b or
Fig.~\ref{hfb}e and Fig.~\ref{hfb}f. The effect is definitely
stronger for the SkO$_{T^\prime}$ than for the SkO, indicating
that the reduction of the spin-orbit combined with strong attractive tensor
fields can affect deformation properties which are rather well captured
by conventional Skyrme forces.  Hence, nuclear deformability  in the
presence of strong attractive tensor fields and reduced SO potential
poses very stringent test for this new class of functionals.

The aim of this section is to show that deformation
properties indeed depend sensitively on the balance between the SO and
tensor fields.  We will discuss two contrasting cases.
First, we will consider an example of the yrast superdeformed (SD) band in
$^{56}$Ni~\cite{[Rud99w]}.  The band is formed by promoting two protons and
two neutrons from the $1f_{7/2}$ to the $1f_{5/2}$ orbital or, in terms
of more appropriate from mean-field point of view asymptotic Nilsson model
quantum numbers, from $[303]7/2$ to $[321]3/2$ Nilsson orbital as illustrated
schematically in Fig.~\ref{56ni}a.

In the spherical ground state of $^{56}$Ni the entire $1f_{7/2}$ is
fully occupied while the $1f_{5/2}$ is empty.  This creates large
spin-asymmetry and, in turn, large contribution to the ground state
due to the tensor field.  By promoting four-particles from the
$1f_{7/2}$ to the $1f_{5/2}$ orbit one creates the SD state which
has reduced spin-asymmetry as compared to the ground state.
The reduced tensor field shifts the SD state up in energy with
respect to the ground state.  The subsequent reduction of the SO potential shifts
the $1f_{5/2}$ orbit and, in turn, the entire SD band down in energy.
This compensating mechanism is illustrated in
Fig.~\ref{56ni}b,c,d.  The figures show the results of the self-consistent
quadrupole-constrained HF calculations for the ground state (GS) and the SD
configurations.  Since it is impossible to go diabatically through the
GS--SD configuration crossing region the self-consistent results
for the GS and SD configurations
were, for the sake of simplicity and clarity, extrapolated diabatically
through this region.  This does not affect the physics discussed below.

Fig.~\ref{56ni}b shows the calculations performed using the
conventional SkO force.  These calculations predict the $0^+$ SD state to be
excited by $\sim$4\,MeV with respect to the GS what agrees
quite well with the empirical estimate~\cite{[Rud99w]}.
Readjustment of the tensor coupling constant in the SkO to the value
characteristic for the SkO$_{T\prime}$ functional (this functional
is called SkO$_{TX}$) shifts the position of the $0^+$ SD state by 4\,MeV up
in energy.  This intermediate step is illustrated in Fig.~\ref{56ni}c.
The change in the tensor field on the
passage from the GS to the SD minimum is shown in Fig.~\ref{56ni}e.
Finally, Fig.~\ref{56ni}d shows the result obtained using
full SkO$_{T\prime}$ functional.  The effects of the reduced SO and
strong attractive tensor fields almost cancel each other
restoring the position of the $0^+$ SD state close to its empirical
(and close to the SkO) value.

\vspace{0.3cm}

\begin{figure}[t]
\includegraphics[width=0.4\textwidth, clip]{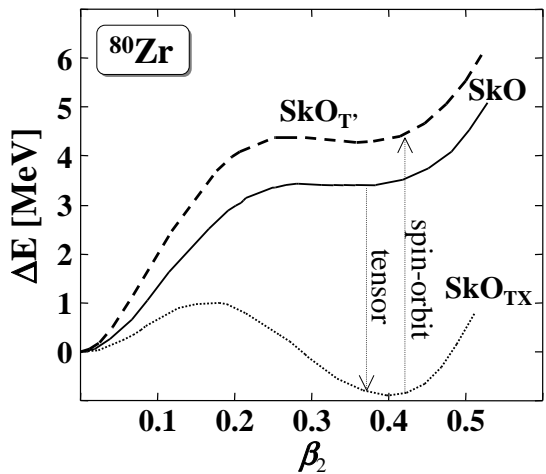}
\caption{Potential energy curves calculated using
quadrupole-constrained Skyrme HFB model.  Solid curve represents
the SkO calculations.  Dotted and dashed curves illustrate
the SkO$_{TX}$ and SkO$_{T^\prime}$ results, respectively.
Note, that tensor field tends to favor
strongly elongated shapes leading to well deformed
absolute minimum.  The subsequent reduction of the SO
strength, however, shifts the entire potential
energy curve up in energy close to its
original SkO position.
}
\label{80zr}
\end{figure}

The second example is shown in Fig.~\ref{80zr}.  The figure
illustrates potential energy curves in $^{80}$Zr calculated using
the HFB  model with volume-delta interaction, $V\delta({\pmb r})$, of
the strength $V$=-190\,MeV in the particle-particle channel.  The three
curves represent the SkO, SkO$_{TX}$, and SkO$_{T^\prime}$
functionals in the particle-hole channel.  All curves are normalized to the
spherical minimum in order to facilitate further discussion.

Unlike the $^{56}$Ni, the $^{80}$Zr is spin-saturated in the spherical
minimum.  Building up deformation is associated in this nucleus with
partial occupation of the $1g_{9/2}$ sub-shell.  It leads to increasing
spin asymmetry and, in turn, to extra attraction due to the tensor terms.
This effect is clearly visible for the SkO$_{TX}$ functional, see
Fig.~\ref{80zr}.  The mechanism is strong enough to create a well deformed minimum.
In this case, reduction of the SO strength shifts the $1g_{9/2}$
sub-shell up with respect to the negative parity $fp$ levels.
In turn, the well deformed minimum is also lifted up in energy
ending up slightly higher as compared to the SkO prediction.

These two examples show that the SkO$_{T^\prime}$ functional have
deformation properties quite similar to the conventional Skyrme functionals
at least in isoscalar $N\approx Z$ nuclei.
The situation is slightly more
intricate in the isovector channel
due to the uncertainties of the $C_1^{\nabla J}$ strength and, in turn, in the
$C_1^{J}$ coupling constant.
Nevertheless, the two cases analyzed above clearly show that:
\begin{itemize}
\item
Nuclear deformation properties strongly depend on the~balance between
tensor and spin-orbit terms

\item
Detailed and systematic studies of nuclear deformation in the~presence
of strong tensor fields open up new, promising venue which may help to
tune the effective tensor coupling constants $C_t^J$.

\item
There is a~large,
and so far completely unexplored, potential to modify deformation properties
by extending the tensor term from the uniform form
used here to the non-uniform form (\ref{nonuni}).

\item
Possible modifications of collective rotational motion, which is inherently
related to the spontaneous breaking of spherical symmetry within the EDF
formalism, opens up yet another almost completely unexplored area in studying
tensor fields.
\end{itemize}

\section{Summary and conclusions}\label{sec7}

The direct fit of the isoscalar spin-orbit
and both isoscalar and isovector tensor coupling constants to
the $f_{5/2}-f_{7/2}$ SO splittings in $^{40}$Ca, $^{56}$Ni, and $^{48}$Ca
requires ({\it i\/}) a drastic reduction of the isoscalar SO strength
and ({\it ii\/}) strong attractive tensor coupling constants~\cite{[Zal08]}.
In this work we  address the global nuclear structure
consequences of this novel fitting strategy of the nuclear EDF.
Among others, we show that contribution to the nuclear
binding energy due to the tensor field shows generic {\it magic structure\/}
with the {\it tensorial
magic numbers\/} at $N(Z)$=14, 32, 56, or 90 corresponding to the maximum
spin-asymmetries in $1d_{5/2}$, $1f_{7/2}\oplus 2p_{3/2}$,  $1g_{9/2}\oplus
2d_{5/2}$ and $1h_{11/2}\oplus 2f_{7/2}$
single-particle configurations and that these numbers are
smeared out by pairing correlations and deformation effects.

We explicitly construct the functional, dubbed SkO$_{T^\prime}$, which
is able to reproduce simultaneously the $f_{5/2}-f_{7/2}$ SO
splittings in $^{40}$Ca, $^{56}$Ni, and $^{48}$Ca nuclei and the binding
energies of spherical nuclei.  In fact, one can construct many
parameterizations reproducing these data in a~more or less equivalent manner.
This is due to the fact that multi-dimensional merit-functions which are
minimized in the fitting process are flat in certain directions and the
entire minimization procedure is not well-defined.
Reasonable performance on the binding energies of the SkO$_{T^\prime}$
functional which is characterized by its non-conventional isovector SO term
is very interesting by itself. Indeed, this result
may open up a possibility to bridge the $C^{J}$ and $C^{\nabla J}$ coupling constants
resulting from (local) fits to the SPE with the values resulting from global
large-scale fits to the binding energies by exploring non-standard local functionals.

Using the SkO$_{T^\prime}$ functional we analyze polarization effects exerted
by one particle and one hole on the values of the $f_{5/2}-f_{7/2}$ SO
splittings.  We show that the polarization effects are large and very
sensitive to the way the functional is set up in the time-odd
channel.  Fits to the Landau parameters are uncertain
due to rather poorly known $p$-wave Landau parameters.  In particular, the use
of the $p$-wave Landau parameters deduced from the Gogny force, which
is advocated in Ref.~\cite{[Zdu05xw]}, leads to strong gauge-symmetry violating
effects and, in turn, large differences in the $C^s$ coupling constants
between the gauge-invariant SkO$_{T^\prime L_S}$ functional
and the gauge-symmetry violating functional SkO$_{T^\prime L_B}$.

We also demonstrate that deformation properties in
atomic nuclei can be easily and strongly modified in the presence
of strong tensor fields and that these properties are extremely
sensitive to the~balance between the tensor and the SO coupling
constants.  We show that, in the~particular case of the SkO$_{T^\prime}$
functional, the tensor effects  are almost perfectly compensated,
at least in the isoscalar channel, by the reduced SO potential.
We suggest that the role of the tensor interaction, in particular
in the time-odd channel, can be studied through dynamical effects
induced by fast nuclear rotation.  For this purpose one needs to
select and use nuclear states representing, as close as possible,
an unperturbed single-particle
motion in order to suppress other effects or correlations which may obscure
the analysis. The examples include superdeformed bands which are known
to be very well described using simple one-body cranking approximation,
see Ref.~\cite{[Sat05]} and references therein, or
terminating states~\cite{[Sat08]}.

\vspace{0.3cm}

This work was supported in part by the Polish Ministry of
Science under Contracts No.~N~N202~328234 and~N~N202~239137, by the
G\"oran Gustafsson Foundation
and by the Swedish Science Research Council (VR).
%\bibliography{C:/Actual/LaTeX/Temp/jd,C:/Actual/LaTeX/Latex.all/jacwit25}

%\bibliographystyle{unsrt}

%\bibliography{jacwit25,tensor}

\end{document}